 \documentclass[preprint,aps]{revtex4}
\usepackage[pdftex]{graphicx}
 
\begin{document}
  \newcommand{\Qed}{\rule{2.5mm}{3mm}}
 \newcommand{\balpha}{\mbox{\boldmath {$\alpha$}}}
 \def\Tr{{\rm Tr}}
 \def\(#1)#2{{\stackrel{#2}{(#1)}}}
 \def\[#1]#2{{\stackrel{#2}{[#1]}}}
 \def\A{{\cal A}}
 \def\B{{\cal B}}
 \def\Sb#1{_{\lower 1.5pt \hbox{$\scriptstyle#1$}}}
 \draft 
\title{Does dark matter consist of baryons of new stable family quarks?\\%
}

\author{G. Bregar, N.S. Manko\v c Bor\v stnik}
\address{Department of Physics, FMF, University of
Ljubljana, Jadranska 19, 1000 Ljubljana}

\date{\today}


\begin{abstract} 
We investigate the possibility that the dark matter consists of 
clusters of the   
heavy family quarks and leptons with   zero Yukawa couplings to the 
lower families. Such a family 
is predicted by the approach unifying spin and charges as the fifth family.   
We make a rough estimation of properties of  baryons  of this new  
family members and study possible limitations on the family properties due to the 
direct experimental and the cosmological evidences. 
\end{abstract}

\maketitle

%
\section{Introduction}
\label{introduction}

Although the origin of the dark matter is 
unknown, its gravitational interaction with the known matter  and other cosmological 
observations 
require from the candidate for the dark matter constituent that: 
i. The scattering amplitude of a cluster of constituents with the ordinary matter 
and among the dark matter clusters themselves must be small enough, 
so that no effect of such scattering has been observed, except possibly 
in the DAMA/NaI~\cite{rita0708} and not (yet?) in the CDMS and other experiments~\cite{cdms}. 
ii. Its  density distribution (obviously different from the ordinary matter density 
distribution) causes that all the stars  within a galaxy rotate approximately with the same
velocity (suggesting that the density is approximately spherically symmetrically distributed, 
descending with the second power of the distance from the center, it is extended also far 
out of the galaxy,  manifesting the gravitational lensing by galaxy clusters).  
iii. The dark matter constituents must be stable in comparison with the age of our universe,  
having obviously for many orders of magnitude different time scale for forming (if at all)  
solid matter than the ordinary matter. 
iv. The dark matter constituents had to be formed during the evolution of our 
universe 
so that they contribute today the main part of the matter ((5-7) times as much as the 
ordinary matter).  


There are several candidates for the massive dark matter constituents in the literature, 
known as WIMPs 
(weakly interacting massive particles), the references can be found in~\cite{dodelson,rita0708}. 
In this paper we discuss the possibility that the dark matter constituents are 
clusters of a stable (from the point of view of the age of the universe)
family of quarks and leptons. Such a family is predicted 
by the approach unifying spin and  charges~\cite{pn06,n92,gmdn07}, proposed by one 
of the authors of this letter: N.S.M.B.. 

The origin of families is  not understood~\footnote{Although the assumptions of the standard model 
of the electroweak and colour interactions leads to predictions with so far  not in disagreement with the 
experimental data, almost all the assumptions wait for explanations, like: 
What is the origin of families?  
Why do only the left handed quarks and leptons carry the weak charge, while the right handed ones 
do not? Why do particles carry the observed $SU(2), U(1)$ and $SU(3)$ charges? Where does the Higgs field 
originate from? And others.} so far. There are several 
attempts in the literature trying to understand the origin of families. All of them, however, 
in one or another way (for example through choices of appropriate groups) simply postulate that there are at 
least three families, as 
does the standard model. 
Proposing the (right) mechanism for generating families is to our understanding 
one of the most promising guides to 
physics beyond the standard model. 
{\it The approach unifying spin and charges is 
offering the mechanism for the appearance of families.} It introduces the second 
kind~\cite{pn06,n92,n93,hn02hn03} 
of the Clifford algebra objects,  which generates families as the  
equivalent representations 
to the Dirac spinor representation~\footnote{The references~\cite{n93,hn02hn03} show that there are 
two (only two) kinds of the Clifford algebra objects, one used by Dirac to describe the spin of fermions. 
The second kind forms the equivalent representations with respect to the Lorents group for spinors~\cite{pn06}. 
The families do form the equivalent representations with repect to the Lorentz group.}. 
The approach predicts from the simple starting action for the fermions, which carry two kinds of the 
Clifford algebra objects' quantum numbers, more than the observed three families. It predicts 
two times  four families with masses several orders of magnitude bellow the 
unification scale of the three observed charges.  
Since due to the approach 
(after assuming a particular way of nonperturbative breaking the starting symmetry) 
the fifth family decouples in the Yukawa couplings from 
the lower four families~\cite{gmdn07}, 
the quarks and the leptons of the fifth family are stable as required by the condition iii.. 
Since the masses of all the members of the fifth family lie much above the known three  
and the predicted fourth family masses 
(the fourth family might according to the first very rough estimates  be even seen at the LHC), 
the baryons made out of the fifth family form small enough  clusters, so that 
 their scattering amplitude among themselves and with the ordinary matter  
is small enough and also the number of clusters is low enough to fulfil the conditions i. and iii..


There are several assessments about masses of a possible (non stable) fourth family of quarks and leptons, 
which follow from the analyses of the existing experimental data and the cosmological observations. 
Although most of physicists have  doubts about the existence of more 
than three families, the analyses clearly show that neither the experimental electroweak data~\cite{okun,pdg}, 
nor the cosmological observations~\cite{pdg} forbid the existence of more than three (so far 
observed) families, as long as the masses of the fourth family quarks are higher than a few hundred 
GeV and the masses of the fourth family leptons higher than one hundred GeV. 
Our stable  fifth family baryons, which might form the dark matter, also   
do not contradict the so far observed experimental data---as it is the measured 
(first family) baryon number and its ratio to the photon  energy density, 
as long as the fifth family quarks are 
 heavy enough ($>$1 TeV). (This is true for any stable heavy family.)
Namely, all the measurements, which connect the baryon and the photon 
energy density, relate to the moment(s) in the history of 
the universe, when the baryons of the first family where formed ($m_1 c^2 \approx 
k_b T \approx 1$ GeV)  and the electrons and nuclei   formed  atoms ($k_b \,T 
 \approx 1$ eV). The chargeless (with respect to the colour and electromagnetic 
 charges and not with respect to the weak charge) clusters of the fifth family were 
 formed long before (at $T k_b \approx E_{c_5}$ (see Table~\ref{TableI.})). They  manifest 
 after decoupling from the plasma (with their small number density and  small cross 
 section) (almost) only their gravitational  interaction.


In this paper we estimate the properties of the fifth family members ($u_5,d_5,\nu_5,e_5$) 
for which the approach unifying spin and charges predicts that they  
have the properties of the lower four families: the same family members with the same charges 
and interacting correspondingly with the same gauge fields. We estimate the masses of the fifth family quarks, 
their behaviour in the evolution of the universe, their formation of 
clusters, properties of these clusters  and the behaviour of clusters 
when scattering among themselves and with the ordinary matter (the first family baryons and leptons). 

We use a simple (the hydrogen-like) model~\cite{gnBled07} to estimate
the size and the binding energy of the fifth family baryons, assuming that the fifth family 
quarks are heavy enough to interact mostly by exchanging one gluon. 
Solving the corresponding Boltzmann equations we estimate the behaviour of quarks and 
anti-quarks of the fifth family during the evolution of our universe, assuming that there is no 
excess of quarks over anti-quarks, concluding that quarks and anti-quarks, which   
succeeded to form neutral (with respect to the colour and electromagnetic charge) clusters,  might 
now form the dark matter, while the  rest disappeared at the colour phase 
transition at around $1$ GeV and lower.  
We also estimate the behaviour of 
our fifth family clusters if hitting the DAMA/NaI---DAMA-LIBRA~\cite{rita0708} and CDMS~\cite{cdms} 
experiments estimating the limitations the DAMA/NaI experiments put on our fifth family quarks when    
recognizing that CDMS has not found any event yet.  

\vspace{1mm}

\section{Properties of clusters of the heavy family}
\label{properties}

Let us  assume that there is a heavy family of quarks and leptons as 
predicted by the approach unifying spin and charges, with masses several orders of 
magnitude greater than those of the known three families,  decoupled in the Yukawa couplings from 
the lower mass families and with 
the charges and their couplings to the gauge fields of the known families (which all seems, 
due to our estimate predictions of the approach, reasonable assumptions).   
Families distinguish among themselves (besides in masses) 
in the family index (in the quantum number, which in the approach is determined  
by the second kind of the Clifford algebra objects operators~\cite{pn06,n92,n93} 
$\tilde{S}^{ab}=\frac{i}{4}(\tilde{\gamma}^a \tilde{\gamma}^b - \tilde{\gamma}^b 
\tilde{\gamma}^a)$, anti-commuting with the Dirac $\gamma^a$'s), and 
(due to the Yukawa couplings)  in their masses.  

For a heavy enough family the properties of baryons (protons $p_5$ $(u_5 u_5 d_5)$, 
neutrons $n_5$ $(u_5 d_5 d_5)$, $\Delta_{5}^{-}$, $\Delta_{5}^{++}$) 
made out of  
quarks $u_5$ and $d_5$ can be estimated by using the non relativistic Bohr-like model 
with the $\frac{1}{r}$ 
dependence of the potential  
between a pair of quarks  $V= - \frac{2}{3} \frac{\hbar c \,\alpha_c}{r}$, where $\alpha_c$ is in this case the 
colour coupling constant. 
Equivalently goes for anti-quarks. 
This is a meaningful approximation as long as the   
one gluon exchange is the dominant contribution to the interaction among quarks,  
that is as long as excitations  of a cluster are not influenced by  the linearly rising 
part of the potential~\footnote{Let us tell that a simple bag model evaluation does not 
contradict such a simple Bohr-like model.}. The electromagnetic 
and weak interaction contributions are  more than $10^{-2}$ times smaller. 
%
Which one of $p_5$, $n_5$, or maybe $\Delta_{5}^-$ or $\Delta_{5}^{++}$,  
is a stable fifth family baryon, depends on the ratio of the bare masses 
$m_{u_5}$ and  $m_{d_5}$, as well as on the  weak and the 
electromagnetic interactions among quarks. 
If $m_{d_5}$ is appropriately 
smaller than $m_{u_5}$ so that the  
weak and electromagnetic interactions favor the neutron $n_5$, then $n_5$ is 
a colour singlet electromagnetic chargeless stable cluster of quarks, with 
the weak charge $-1/2$. 
If $m_{d_5}$ is larger (enough, due to the stronger electromagnetic repulsion among 
the two $u_5$ than among the two $d_5$) than $m_{u_5}$, the proton $p_5$ which is 
a colour singlet stable nucleon with the weak charge $1/2$,  
needs the electron $e_5$ or $e_1$ to form  a stable  electromagnetic 
chargeless cluster.  
An atom made out of only fifth family members might be lighter or not than $n_5$, 
depending on the masses of the fifth family members. 

Neutral (with respect to the electromagnetic and colour charge) 
particles that constitute the dark matter can be $n_5,\nu_5$ 
or  charged baryons like  $p_5, \Delta^{++}_5$, $\Delta^{-}_5$, forming neutral atoms with
$e^{-}_5$ or $e^{+}_5$, correspondingly. We treat the case that $n_5$ as well as $\bar{n}_5$ 
form the major part of the dark matter, assuming 
that $n_5$ (and $\bar{n}_5$) are stable baryons (anti-baryons). Taking  $m_{\nu_5}< m_{e_5}$ 
also $\nu_5$ contributes to the dark matter.

In the Bohr-like model 
we obtain if neglecting more than one gluon exchange contribution
%
\begin{eqnarray}
\label{bohr}
E_{c_{5}}\approx -3\; \frac{1}{2}\; \left( \frac{2}{3}\, \alpha_c \right)^2\; \frac{m_{q_5}}{2} c^2,
\quad r_{c_{5}} \approx  \frac{\hbar c}{ \frac{2}{3}\;\alpha_c \frac{m_{q_5}}{2} c^2}. 
\end{eqnarray}
The mass of the cluster is approximately $m_{c_5}\, c^2 \approx  
3 m_{q_5}\, c^2(1- (\frac{1}{3}\, \alpha_c)^2)$. We use the  factor of $\frac{2}{3}$ 
for a  two quark pair potential and of $\frac{4}{3}$ for a quark and an anti-quark pair potential. 
If treating correctly the three quarks' (or anti-quarks') center of mass motion in the 
hydrogen-like model, allowing  the hydrogen-like functions to adapt the width as presented in 
Appendix~I,
the factor $-3\; \frac{1}{2}\; (\frac{2}{3})^2\; \frac{1}{2}$ in Eq.~\ref{bohr} is replaced by
$0.66$, and the mass of the cluster is accordingly $3 m_{q_5} c^2(1-0.22\, \alpha_{c}^2)$, while 
the average radius takes the values as presented in  Table~\ref{TableI.}.

Assuming that the coupling constant   
of the colour charge  $\alpha_c$   runs with the kinetic energy $- E_{c_{5}}/3$ and taking into account 
the number of families which contribute to the running coupling constant in dependence on the kinetic energy 
(and correspondingly on the mass of the fifth family quarks)
%
%
we estimate  the properties of a baryon as presented on Table~\ref{TableI.} (the table 
is calculated from the hydrogen-like model presented in Appendix~I), 
\begin{table}
\begin{center}  
\begin{tabular}{||c||c|c|c|c|}
\hline
$\frac{m_{q_5} c^2}{{\rm TeV}}$ & $\alpha_c$ & $\frac{E_{c_5}}{m_{q_5} c^2}$ & 
$\frac{r_{c_5}}{10^{-6}{\rm fm}}$ & $\frac{\Delta m_{ud} c^2}{{\rm GeV}}$ 
\\
\hline
\hline
$1   $ & 0.16   & -0.016   & $3.2\, \cdot 10^3$   & 0.05		             \\
\hline
$10  $ & 0.12   & -0.009   & $4.2\, \cdot 10^2$   & 0.5           \\
\hline
$10^2$ & 0.10   & -0.006   & $52$            & 5           \\
\hline
$10^3$ & 0.08   & -0.004   & $6.0$           & 50           \\
\hline
$10^4$ & 0.07   & -0.003   & $0.7$           & $5 \cdot 10^2$           \\
\hline
$10^5$ & 0.06   & -0.003   & $0.08$          & $5 \cdot 10^3$            \\
\hline
\end{tabular}
\end{center}
\caption{\label{TableI.} 
The properties of a cluster of the fifth family quarks
within the extended Bohr-like (hydrogen-like) model from Appendix~I. 
$m_{q_5}$ in TeV/c$^2$ is the assumed fifth family quark mass,
$\alpha_c$ is the coupling constant 
of the colour interaction at $E\approx (- E_{c_{5}}/3)\;$ (Eq.\ref{bohr})   
which is the kinetic energy 
of  quarks in the baryon, 
$r_{c5}$ is the corresponding average  radius. Then  $\sigma_{c_5}=\pi r_{c_5}^2 $  
is the corresponding scattering cross section.} 
\end{table}

The binding energy is approximately  $\frac{1}{ 100}$  of the mass 
of the cluster (it is $\approx \frac{\alpha_{c}^2}{3}$).  The baryon $n_5$ ($u_{5} d_{5} d_{5}$) 
is lighter than the baryon $p_{5}$,   ($u_{q_5} d_{q_5} d_{q_5}$) 
if $\Delta m_{ud}=(m_{u_5}-m_{d_5})$ is smaller than $(0.05,0.5,5, 50, 500, 5000)$ GeV  
for the six  values of the $m_{q_5} c^2$ on Table~\ref{TableI.}, respectively. 
We see  from Table~\ref{TableI.} that the ''nucleon-nucleon'' 
force among the fifth family baryons leads to many orders of 
magnitude smaller cross section than in the case 
of the first family nucleons ($\sigma_{c_5}= \pi r_{c_5}^2$ is from $10^{-5}\,{\rm fm}^2$  for 
$m_{q_5} c^2 = 1$ TeV to $10^{-14}\, {\rm fm}^2$  for $m_{q_5} c^2 = 10^5$ TeV). 
Accordingly is the scattering cross section between two  fifth family baryons   
determined by the weak interaction as soon as the mass   exceeds  several GeV.

If a cluster of the heavy (fifth family) quarks and leptons and  of the 
ordinary (the lightest) family is made, 
then, since ordinary family   dictates the radius and the excitation energies  
of a cluster, its 
properties are not far from the properties of the ordinary hadrons and atoms, except that such a  
cluster has the mass dictated by the heavy family members. 
\section{Dynamics of a heavy family baryons in our galaxy}
\label{dynamics}

%

There are experiments~\cite{rita0708,cdms} which are trying to directly measure the dark matter clusters. Let us 
make a short introduction into these measurements. We shall treat our fifth family clusters in particular.
The density of the dark 
matter $\rho_{dm}$ in the Milky way can be evaluated from the measured rotation velocity  
of  stars and gas in our galaxy, which appears to be approximately independent of the distance $r$ from the 
center of our galaxy. For our Sun this velocity 
is $v_S \approx (170 - 270)$ km/s. $\rho_{dm}$ is approximately spherically symmetric distributed 
and proportional to $\frac{1}{r^2}$.  Locally (at the position of our Sun) $\rho_{dm}$ 
is known within a factor of 10 to be 
$\rho_0 \approx 0.3 \,{\rm GeV} /(c^2 \,{\rm cm}^3)$, 
we put $\rho_{dm}= \rho_0\, \varepsilon_{\rho},$ 
with $\frac{1}{3} < \varepsilon_{\rho} < 3$. 
The local velocity distribution of the dark matter cluster $\vec{v}_{dm\, i}$, in the 
velocity  class 
$i$ of clusters, can only be estimated, 
results depend strongly on the model. Let us illustrate this dependence.  
In a simple model that all the clusters at any radius $r$ from the center 
of our galaxy travel in all possible circles around the center so that the paths are 
spherically symmetrically distributed, the velocity of a cluster at the position of 
the Earth is equal to $v_{S}$, the velocity of our Sun in the absolute value,
but has all possible orientations perpendicular to the radius $r$ with  equal probability.
In the model 
that the clusters only oscillate through the center of the galaxy, 
the velocities of the dark matter clusters at the Earth position have values from 
zero to the escape velocity, each one weighted so that all the contributions give  
$ \rho_{dm} $. 
Many other possibilities are presented in the references cited in~\cite{rita0708}. 

The velocity of the Earth around the center of the galaxy is equal to:  
$\vec{v}_{E}= \vec{v}_{S} + \vec{v}_{ES} $, with $v_{ES}= 30$ km/s and 
$\frac{\vec{v}_{S}\cdot \vec{v}_{ES}}{v_S v_{ES}}\approx \cos \theta \, \sin \omega t, \theta = 60^0$. 
Then the velocity with which the dark matter cluster of the $i$- th  velocity class  
hits the Earth is equal to:  
$\vec{v}_{dmE\,i}= \vec{v}_{dm\,i} - \vec{v}_{E}$. 
$\omega $ 
determines the rotation of our Earth around the Sun.

One finds for the flux 
of the  
dark matter clusters hitting the Earth:    
$\Phi_{dm} = \sum_i \,\frac{\rho_{dm \,i}}{m_{c_5}}  \,
|\vec{v}_{dm \,i} - \vec{v}_{E}|  $ 
to be approximately  (as long as $\frac{v_{ES}}{|\vec{v}_{dm \,i}- \vec{v}_S|}$ is small
) equal to  
\begin{eqnarray}
\label{flux}
\Phi_{dm}\approx \sum_i \,\frac{\rho_{dm \,i}}{m_{c_5}}  \,
\{|\vec{v}_{dm \,i} - \vec{v}_{S}| - \vec{v}_{ES} \cdot \frac{\vec{v}_{dm\, i}- \vec{v}_S}{
|\vec{v}_{dm \,i}- \vec{v}_S|} \}.
\end{eqnarray}
Further terms are neglected. 
We shall approximately take that
$\sum_i \, |\vec{v_{dm \,i}}- \vec{v_S}| \,\rho_{dm \,i} \approx \varepsilon_{v_{dmS}} 
\, \varepsilon_{\rho}\,  v_S\, \rho_0 $, 
and correspondingly 
$ \sum_i \, \vec{v}_{ES}  \cdot \frac{\vec{v}_{dm \,i}- \vec{v}_S}{
|\vec{v}_{dm \,i}- \vec{v}_S|} \approx v_{ES} \varepsilon_{v_{dmS}}
\cos \theta \, \sin \omega t $, 
(determining the annual modulations observed by DAMA~\cite{rita0708}). 
Here $\frac{1}{3} < \varepsilon_{v_{dmS}} < 
3$ and $\frac{1}{3} < \frac{\varepsilon_{v_{dmES}}}{\varepsilon_{v_{dmS}}} < 3$ are
estimated with respect to experimental and (our) theoretical evaluations.  

Let us evaluate the cross section for our heavy dark matter baryon to elastically
(the excited states of nuclei,  
which we shall treat, I and Ge, are at $\approx 50$ keV 
or higher and are very narrow, while the average recoil energy of Iodine is expected to be 
$30$ keV) 
scatter  on an ordinary nucleus with $A$ nucleons 
$\sigma_{A} = 
\frac{1}{\pi \hbar^2} <|M_{c_5 A}|>^2 \, m_{A}^2$. 
For our heavy dark matter cluster 
is  $m_{A}  $  approximately the mass of the ordinary nucleus. In the case of a 
coherent scattering (if recognizing that $\lambda= \frac{h}{p_A}$ is for a nucleus large enough 
to make scattering coherent when the mass of  the cluster is 
 $1$ TeV or more and its velocity 
$\approx v_{S}$), the cross section is  almost independent of the recoil 
velocity of the nucleus. 
For the case that the ''nuclear force'' as manifesting  in the cross section $\pi\, (r_{c_5})^2$ 
in Eq.(\ref{bohr}) 
brings the main contribution~\footnote{The very heavy colourless cluster of three quarks,  
hitting with the relative velocity $\approx 200 \,{\rm km}/{\rm s}$ the nucleus of the first 
family quarks, ''sees'' the (light) quark  $q_1$ of the 
nucleus through the cross section $\pi\, (r_{c_5})^2$.
But since the quark $q_{1}$ is at these velocities strongly bound to the proton and the 
proton to the nucleus,  the hole nucleus takes the momentum.} 
the cross section  is  proportional to $(3A)^2$ 
(due to the square of the matrix element) times $(A)^2$ (due to the mass of the nuclei 
$m_A\approx 3 A \,m_{q_1}$, with $m_{q_1}\, c^2 \approx \frac{1 {\rm GeV}}{3}$).  
When $m_{q_5}$ is  heavier than $10^4 \, {\rm TeV}/c^2$ (Table~\ref{TableI.}), 
the weak interaction dominates and $\sigma_{A}$ is proportional to $(A-Z)^2 \, A^2$, 
since to $Z^0$ boson exchange only neutron gives an appreciable contribution. 
Accordingly we have,  when the ''nuclear force'' dominates,
$\sigma_A \approx \sigma_{0} \, A^4 \, \varepsilon_{\sigma}$, with 
$\sigma_{0}\, \varepsilon_{\sigma}$, which is $\pi r_{c_5}^2 \, 
\varepsilon_{\sigma_{nucl}} $ and with 
 $\frac{1}{30} < \varepsilon_{\sigma_{nucl}} < 30$.  
$\varepsilon_{\sigma_{nucl}}$
takes into account the roughness 
with which we treat our  heavy baryon's properties and the scattering procedure.  
When the weak interaction dominates, $ \varepsilon_{\sigma}$ is smaller and we have $  
 \sigma_{0}\, \varepsilon_{\sigma}=(\frac{m_{n_1} G_F}{\sqrt{2 \pi}} 
\frac{A-Z}{A})^2 \,\varepsilon_{\sigma_{weak}}  $
($=( 10^{-6} \,\frac{A-Z}{ A} \, {\rm fm} )^2 \,\varepsilon_{\sigma_{weak}} $), 
$ \frac{1}{10}\, <\,  \varepsilon_{\sigma_{weak}} \,< 1$. The weak force is pretty accurately 
 evaluated, but the way how we are averaging is not.

\section{Direct measurements of the fifth family  baryons as dark matter constituents} 
\label{directmeasurments}

We are making very rough estimations of what the  
 DAMA~\cite{rita0708} and CDMS~\cite{cdms} experiments are measuring, provided that the 
 dark matter clusters are made out 
 of our (any) heavy family quarks as discussed above. 
 We are looking for limitations these two experiments might put on 
 properties of our heavy family members. 
 We discussed about our estimations and their relations to the measurements 
 with R. Bernabei~\cite{privatecommRBJF} and 
 J. Filippini~\cite{privatecommRBJF}. 
 Both pointed out (R.B. in particular) that the two experiments can hardly be compared, 
 and that our very approximate estimations may be right only within the orders of magnitude. 
 We are completely aware of how rough our estimation is, 
 yet we conclude that, since the number of measured events  is  proportional to 
 $(m_{c_5})^{-3}$ 
 for masses $\approx 10^4$ TeV or smaller (while for 
 higher masses, when the weak interaction dominates, it is proportional to  
 $(m_{c_5})^{-1}$) that even such rough  estimations   
 may in the case of our heavy baryons say whether both experiments
 do at all measure our (any) heavy family clusters, if one experiment 
 clearly sees  the dark matter signals and the 
 other does not (yet?) and we accordingly estimate the mass of our cluster. 
 
 Let $N_A$ be the number of nuclei of a type $A$ in the 
 apparatus  
 (of either DAMA~\cite{rita0708}, which has $4\cdot 10^{24}$ nuclei per kg of $I$, 
 with $A_I=127$,  
  and  $Na$, with $A_{Na}= 23$ (we shall neglect $Na$), 
 or of CDMS~\cite{cdms}, which has $8.3 \cdot 10^{24}$ of $Ge$ nuclei 
 per kg,  with $A_{Ge}\approx 73$). 
 At velocities  of a dark matter cluster  $v_{dmE}$ $\approx$ $200$ km/s  
 are the $3A$ scatterers strongly bound in the nucleus,    
 so that the whole nucleus with $A$ nucleons elastically scatters on a 
 heavy dark matter cluster.  
Then the number of events per second  ($R_A$) taking place 
in $N_A$ nuclei   is  due to the flux $\Phi_{dm}$ and the recognition that the cross section 
is at these energies almost independent 
of the velocity 
equal to
\begin{eqnarray}
\label{ra}
R_A = \, N_A \,  \frac{\rho_{0}}{m_{c_5}} \;
\sigma(A) \, v_S \, \varepsilon_{v_{dmS}}\, \varepsilon_{\rho} \, ( 1 + 
\frac{\varepsilon_{v_{dmES}}}{\varepsilon_{v_{dmS}}} \, \frac{v_{ES}}{v_S}\, \cos \theta
\, \sin \omega t).
\end{eqnarray}
Let $\Delta R_A$ mean the amplitude of the annual modulation of $R_A$ 
\begin{eqnarray}
\label{anmod}
\Delta R_A &=& R_A(\omega t = \frac{\pi}{2}) - R_A(\omega t = 0) = N_A \, R_0 \, A^4\, 
\frac{\varepsilon_{v_{dmES}}}{\varepsilon_{v_{dmS}}}\, \frac{v_{ES}}{v_S}\, \cos \theta,
\end{eqnarray}
where $ R_0 = \sigma_{0} \, \frac{\rho_0}{m_{c_5}} \,  v_S\, \varepsilon$, 
$R_0$ is for the case that the ''nuclear force'' 
dominates $R_0 \approx  \pi\, (\frac{3\, \hbar\, c}{\alpha_c \, m_{q_5}\, c^2})^2\, 
\frac{\rho_0}{m_{q_5}} \, v_S\, \varepsilon$, with 
$\varepsilon = 
\varepsilon_{\rho} \, \varepsilon_{v_{dmES}} \varepsilon_{\sigma_{nucl}} $.  $R_0$ is therefore 
proportional to $m_{q_5}^{-3}$. 
We estimated  $10^{-3} < \varepsilon < 10^3$,   
which demonstrates both, the uncertainties in the knowledge about the dark matter dynamics 
in our galaxy and our approximate treating of the dark matter properties.  
(When for $m_{q_5} \, c^2 > 10^4$ TeV the weak interaction determines the cross section  
$R_0 $ is in this case proportional to $m_{q_5}^{-1}$.) 
We estimate that an experiment with $N_A$ scatterers  should  measure the amplitude
$R_A \varepsilon_{cut\, A}$, with $\varepsilon_{cut \, A}$ determining  the efficiency  of 
a particular experiment to detect a dark matter cluster collision. 
For small enough $\frac{\varepsilon_{v_{dmES}}}{\varepsilon_{v_{dmS}}}\, 
\frac{v_{ES}}{v_S}\, \cos \theta$ we have 
\begin{eqnarray}
R_A \, \varepsilon_{cut \, A}  \approx  N_{A}\, R_0\, A^4\, 
 \varepsilon_{cut\, A} = \Delta R_A \varepsilon_{cut\, A} \,
 \frac{\varepsilon_{v_{dmS}}}{\varepsilon_{v_{dmES}}} \, \frac{v_{S}}{v_{ES}\, \cos \theta}. 
\label{measure}
\end{eqnarray}
If DAMA~\cite{rita0708}   is measuring 
our  heavy  family baryons 
 scattering mostly on $I$ (we neglect the same number of $Na$,  with $A =23$),  
then the average $R_I$ is 
\begin{eqnarray}
\label{ridama}
R_{I} \varepsilon_{cut\, dama} \approx  \Delta R_{dama} 
\frac{\varepsilon_{v_{dmS}}}{\varepsilon_{v_{dmES}}}\,
\frac{v_{S}  }{v_{ES}\, \cos 60^0 } ,
\end{eqnarray}
with $\Delta R_{dama}\approx 
\Delta R_{I}  \, \varepsilon_{cut\, dama}$, this is what we read from their papers~\cite{rita0708}.  
In this rough estimation 
most of unknowns about the dark matter properties, except the local velocity of our Sun,  
the cut off procedure ($\varepsilon_{cut\, dama}$) and 
$\frac{\varepsilon_{v_{dmS}}}{\varepsilon_{v_{dmES}}}$,
(estimated to be $\frac{1}{3} < \frac{\varepsilon_{v_{dmS}}}{\varepsilon_{v_{dmES}}} < 3$), 
 are hidden in $\Delta R_{dama}$. If we assume that the 
Sun's velocity is 
$v_{S}=100, 170, 220, 270$ km/s,  we find   $\frac{v_S}{v_{ES} \cos \theta}= 7,10,14,18, $ 
respectively. (The recoil energy of the nucleus $A=I$ changes correspondingly 
with the square of   $v_S $.)
DAMA/NaI, DAMA/LIBRA~\cite{rita0708} publishes 
$ \Delta R_{dama}= 0.052  $ counts per day and per kg of NaI. 
Correspondingly  is $R_I \, \varepsilon_{cut\, dama}  = 
 0,052 \, \frac{\varepsilon_{v_{dmS}}}{\varepsilon_{v_{dmES}}}\, \frac{v_S}{v_{SE} \cos \theta} $ 
counts per day and per kg. 
CDMS should then in $121$ days with 1 kg of Ge ($A=73$) detect   
$R_{Ge}\, \varepsilon_{cut\, cdms}$
$\approx \frac{8.3}{4.0} \, 
 (\frac{73}{127})^4 \; \frac{\varepsilon_{cut\,cdms}}{\varepsilon_{cut \,dama}}\, 
 \frac{\varepsilon_{v_{dmS}}}{\varepsilon_{v_{dmES}}}\;
 \frac{v_S}{v_{SE} \cos \theta} \;  0.052 \cdot 
 121 \;$ events, 
which is for the above measured velocities equal to $(10,16,21,25)
\, \frac{\varepsilon_{cut\, cdms}}{\varepsilon_{cut\,dama}}\;
\frac{\varepsilon_{v_{dmS}}}{\varepsilon_{v_{dmES}}}$. CDMS~\cite{cdms} 
has found no event.

The approximations we made might cause that the expected  numbers 
$(10,16,21,25)$ multiplied by $\frac{\varepsilon_{cut\,Ge}}{\varepsilon_{cut\,I}}\;
\frac{\varepsilon_{v_{dmS}}}{\varepsilon_{v_{dmES}}}$  
are too high (or too low!!) for a factor let us say $4$ or $10$. 
If in the near future  
CDMS (or some other experiment) 
will measure the above predicted events, then there might be  heavy 
family clusters which form the dark matter. In this case the DAMA experiment   
puts the limit on our heavy family masses (Eq.(\ref{measure})). 

Taking into account all the uncertainties   mentioned above, with the uncertainty with
the ''nuclear force'' cross section included (we evaluate these uncertainties  to be 
$10^{-4}  <\,\varepsilon^{"}\,< 3\cdot 10^3$), we can estimate the mass range of the fifth family quarks 
from the DAMA experiments:  $(m_{q_5}\, c^2)^3= \frac{1}{\Delta R_{dama}} 
N_I\,A^4\, \pi \,(\frac{3 \,\hbar c}{\alpha_c})^2 \,
\rho_0\, c^2\, v_{ES} \,\cos \theta\, \varepsilon^{"}= (0.3\, \cdot 10^7)^3 \, 
\varepsilon^{"} (\frac{0.1}{\alpha_c})^{2}
$ GeV. The lower mass limit, which follows from the DAMA experiment,  is accordingly  
$m_{q_5}\, c^2> 200$ TeV. 
Observing that 
for $m_{q_5} \, c^2> 10^4$ TeV 
the weak force starts to dominate, we estimate the upper limit $m_{q_5}\, c^2< 10^5$ TeV. 
Then
$200 {\rm\; TeV} < m_{q_5} \, c^2 < 10^5$ TeV.

\section{Evolution of the abundance of the fifth family members in the universe}
\label{evolution}

To estimate the behaviour of our stable heavy family quarks and anti-quarks in the expanding 
universe we need to know: \\ 
i.) the masses of our fifth family members, \\
ii.) their particle---anti-particle asymmetry, \\
iii.) their thermally averaged scattering cross sections (as the function of the temperature)   
for scattering    
$\;$ iii.a.) into all the 
relativistic quarks and anti-quarks of lower  families ($<\sigma v>_{q\bar{q}}$),  
$\;$ iii.b.)  into gluons ($<\sigma v>_{gg}$), 
$\;$ iii.c.)  into (annihilating) bound states of a fifth family quark and an anti-quark 
($<\sigma v>_{(q\bar{q})_b}$),  
$\;$ iii.d.) into bound states of two fifth family quarks and into the fifth family baryons 
($<\sigma v>_{c_5}$) 
(and equivalently into two anti-quarks and 
into anti-baryons),\\ 
iv.) the probability for quarks and anti-quarks  of the fifth family to annihilate  
at the colour phase transition ($ T k_b \approx 1$ GeV). 

The quarks and anti-quarks start to freeze out when the temperature of the plasma falls close to  
$m_{q_5}\,c^2/k_b $ ($k_b$ is the Boltzmann constant). They are forming clusters (bound states) 
when the temperature 
falls close to  the binding energy. When the three quarks or three anti-quarks of the 
fifth family form a colourless baryon (or anti-baryon), they decouple from the rest of the plasma due to small 
scattering cross section manifested by the average radius presented in Table~\ref{TableI.}. 

We assume  in this paper  that there is no asymmetry between  quarks and  anti-quarks 
of the fifth family. One evaluates that at  the colour phase transition ($ T k_b \approx 1 {\rm GeV}$)   
the ratio of the scattering time between two coloured quarks (of any kind) and the Hubble time 
is of the order of $\approx 10^{-18}$. Accordingly, although the number of the fifth family 
quarks and anti-quarks is  of the order of $10^{-13}$ smaller than the number of the quarks and anti-quarks 
of the rest of families (as show the solutions of the Boltzmann equations 
presented bellow),  the fifth family quarks and anti-quarks have  enough opportunity 
during the expansion time  (from $10^{-7}$ s to a few seconds) to deplete completely. 
The same would happen to all the lower 
families' quarks and anti-quarks (going due to the Yukawa couplings to the first family members), 
if there would be no   quark---anti-quark asymmetry.  

To see how many fifth family quarks and anti-quarks succeed to form the fifth family 
baryons and anti-baryons we must solve the 
Boltzmann equations as  a function of time (or temperature). Since we do not know the mass of the fifth 
family members (the estimations from the approach unifying spin and charges predict that it must be 
higher than a few TeV and lower than,  say, a few $10^5$ TeV), we take it  
as a parameter. The interaction due to one gluon exchange  is dominant among so massive 
fifth family members and it is at the same time  also much 
larger than the weak and the electromagnetic interaction. Since the one gluon exchange is 
(up to the group properties 
and the coupling constants) equivalent to the one photon exchange, we use for the cross sections (cited above) 
the equivalent cross sections from the electromagnetic case. 
The fifth family quark mass follows from comparing the calculated fifth family 
baryon and anti-baryon number density multiplied by the mass of the clusters 
with the today's dark matter density.

We follow (as much as possible), when estimating the number density of the fifth family quarks $n_{q_5}$ and 
anti-quarks $n_{\bar{q}_5}$ clustered into baryons (with the number density $n_{c_5}$) and anti-baryons 
($n_{\bar{c}_5}$), which to our prediction form  the dark matter today, the ref.~\cite{dodelson}, chapter 3. 
$n_{q_{5}}$ is the number density of all the  fifth family quarks of any colour and spin and 
correspondingly is assumed for the other number densities. 
The following cross sections are needed  in the Boltzmann equations 
\begin{eqnarray}
\label{sigmasq}
< \sigma v>_{q\bar{q}} &=&  \frac{16 \,\pi}{9} 
\;\left( \frac{\alpha_{c} \hbar c^2}{m_{q_{5}}\,c^2}\right)^2 \, c ,\nonumber\\
< \sigma v>_{gg} &=&  \frac{37 \,\pi}{108}\;\left( \frac{\alpha_{c} 
\hbar c^2}{m_{q_{5}}\,c^2}\right)^2\, c, \nonumber\\
< \sigma v>_{c_5} &=& \eta_{c_5}\; 10 \;\left( \frac{\alpha_{c} \hbar c}{m_{g_5}\,c^2} \right)^2\, c\; 
\sqrt{\frac{ E_{c_5}}{T k_b}} \ln{\frac{E_{c_5}}{T k_b}}, \nonumber\\
<\sigma  v>_{(q \bar{q})_b}&=& \eta_{(q \bar{q})_b} \;10 \;\left(
\frac{\alpha_{c} \hbar c}{m_{g_5}\,c^2}\right)^2\, c\; 
\sqrt{\frac{ E_{c_5}}{T k_b}} \ln{\frac{E_{c_5}}{T k_b}}, \nonumber\\
\sigma_{T } &=&  \frac{8 \pi}{3} \left(\frac{\alpha_{c} \hbar c }{m_{g_5} \, c^2}\right)^2,
\end{eqnarray}
where $v$ is the relative velocity between the fifth family  quark and its anti-quark, or between two quarks,
$E_{c_5}$ is the binding energy for a cluster (Eq.~\ref{bohr}). 
$< \sigma v> \;$ is the thermally averaged scattering cross section 
times the  relative velocity: 
i. $< \sigma v>_{q\bar{q}} $ for all the pairs  
of the fifth family  quarks and anti-quarks into all the lower mass (of the four families') 
quarks and anti-quarks, which are, while scattering takes place, ultra relativistic. 
ii. $< \sigma v>_{gg}$  for scattering into gluons.  
iii. $ <\sigma v>_{c_5}$ for two quarks (or two anti-quarks) to scatter into a bound state of 
two quarks (anti-quarks) and from two to three quarks (anti-quarks) colourless clusters. 
We use the equivalent expression as for scattering of an electron and a proton into 
the bound  state of a hydrogen. The parameter $\eta_{c_5}$ takes care of scattering 
of two quarks (anti-quarks) into three 
colourless quarks (or anti-quarks), which are the fifth family baryons (anti-baryons). 
iv. $<\sigma v>_{(q\bar{q})_b}$ for  scattering into a bound state of the fifth family quark and anti-quark,  
which annihilate in the time $\tau_{(q\bar{q})_b} < 10^{-28}$ s. 
$\eta_{(q \bar{q})_b}$ takes care of the roughness of the used formula. 

$\sigma_{T }$ is the Thompson-like scattering cross section of gluons on quarks (or anti-quarks),  
responsible for destroying the bound states of baryons. 

Let $T_0$ be the today's black body radiation temperature, $T(t)$ the actual (studied) temperature, 
$a^2(T^0) =1$ and $a^2(T)= a^2(T(t))$ is the metric tensor component in 
the expanding flat universe---the Friedman-Robertson-Walker metric:  
${\rm diag}\,  g_{\mu \nu} = 
(1, - a(t)^2, - a(t)^2, - a(t)^2),\;$  $(\frac{\dot{a}}{a})^2= \frac{8 \pi G}{3} \rho$, 
with $\rho= \frac{\pi^2}{15} \, g^*\, T^4$, $ \, T=T(t)$,  
$g^*$  measures the number of degrees of freedom of those of the  
four family members (f) and  gauge bosons (b), which are at the treated temperature $T$ 
ultra-relativistic ($g^*= \sum_{i\in {\rm b}} \,g_i + \frac{7}{8} \sum_{i\in {\rm f}} \,g_i$). 
$H_0 \, \approx 1.5\, 10^{-42} \,\frac{{\rm GeV} c}{\hbar c} $ is the present Hubble constant 
and $G = \frac{\hbar c }{ (m_{pl}^2)}$, $m_{pl} c^2 \approx  
1.2 \cdot 10^{19}$ GeV. 

Let us write down the Boltzmann equation, which treats in the expanding universe 
the number density of all the fifth 
family quarks as a function of time $t$. The fifth family quarks scatter with  anti-quark into 
all the other relativistic quarks and anti-quarks ($< \sigma v>_{q\bar{q}}$) and into gluons 
($< \sigma v>_{gg}$). At the beginning, when the quarks are becoming non-relativistic and   
start to freeze out, the  formation of bound states is negligible. One finds~\cite{dodelson} 
the Boltzmann equation for the fifth family quarks $n_{q_5}$ (and  equivalently  for 
anti-quarks $n_{\bar{q}_5}$)
\begin{eqnarray}
\label{boltzq1}
a^{-3}\frac{d( a^3 n_{q_5})}{dt} &=& < \sigma v>_{q\bar{q}}\; n^{(0)}_{q_5} n^{(0)}_{\bar{q}_5}\,
\left( - \frac{n_{q_5} n_{\bar{q}_5}}{n^{(0)}_{q_5} n^{(0)}_{\bar{q}_5}} + 
 \frac{n_{q} n_{\bar{q}}}{n^{(0)}_{q} n^{(0)}_{\bar{q}}} \right) + \nonumber\\ 
&&< \sigma v>_{gg} \; 
n^{(0)}_{q_5} n^{(0)}_{\bar{q}_5}\,
\left( - \frac{n_{q_5} n_{\bar{q}_5}}{n^{(0)}_{q_5} n^{(0)}_{\bar{q}_5}} +  
\frac{n_{g} n_{g}}{n^{(0)}_{g} n^{(0)}_{g}} \right). 
\end{eqnarray}
Let us tell that $n^{(0)}_{i} = g_i\, (\frac{m_i c^2 T k_b}{(\hbar c)^2})^{\frac{3}{2}} 
e^{-\frac{m_i c^2}{T k_b}}$ for 
$m_i c^2 >> T k_b$  
and to  $\frac{g_i}{\pi^2}\, (\frac{T k_b}{\hbar c})^3$ 
for $m_i c^2 << T k_b$.
Since the ultra-relativistic quarks and anti-quarks of the lower families are in the  
thermal equilibrium with the plasma and so 
are gluons, it follows $\frac{n_{q} n_{\bar{q}}}{n^{(0)}_{q} n^{(0)}_{\bar{q}}}=1= 
\frac{n_{g} n_{g}}{n^{(0)}_{g} n^{(0)}_{g}}$. Taking into account that $(a\, T)^3 \, g^*(T)$ is a constant  
it is appropriate~\cite{dodelson} to introduce a new parameter $x=\frac{m_{q_5}c^2}{k_b T}$  and 
the quantity $Y_{q_5}= 
n_{q_5}\, (\frac{\hbar c}{k_b T})^3$, $Y^{(0)}_{q_5}= 
n^{(0)}_{q_5}\, (\frac{\hbar c}{k_b T})^3$.  When taking into account that  the number of 
quarks is the same as the number of anti-quarks, and that 
$\frac{dx}{dt} = \frac{h_m \,m_{q_5}c^2}{x} $, with $h_m = \sqrt{\frac{4 \pi^3 g^*}{45}}\, 
\frac{c}{\hbar c \, m_{pl} c^2}$, Eq.~\ref{boltzq1} transforms into $\frac{dY_{q_5}}{dx} = 
\frac{\lambda_{q_5}}{x^2}\, (Y^{(0)2}_{q_5} - Y^{2}_{q_5}), $ with $\lambda_{q_5} = \frac{(<\sigma v>_{q\bar{q}} + 
<\sigma v>_{gg}) \, m_{q_5} c^2}{h_{m}\, (\hbar c)^3}$. It is this equation which  we are solving 
(up to the region of $x$ when the clusters of quarks 
and anti-quarks start to be formed) to see 
the behaviour of the fifth family quarks as a function of the temperature.  

When the temperature of the expanding universe falls close enough to the binding energy of the cluster 
of the fifth family quarks (and anti-quarks), the bound states of quarks (and anti-quarks) and the 
clusters of fifth family baryons (in our case neutrons $n_{5}$) (and anti-baryons $\bar{n}_{5}$---anti-neutrons) 
start to form.  
The corresponding Boltzmann equation for the number of baryons $n_{c_5}$ reads
\begin{eqnarray}
\label{boltzc}
a^{-3}\frac{d( a^3 n_{c_5})}{dt} &=& < \sigma v>_{c_5}\; n^{(0)^2}_{q_5}\,
\left( \left( \frac{n_{q_5}}{n^{(0)}_{q_5}} \right)^2 -  
\frac{n_{c_5}}{ n^{(0)}_{c_5}} \right).
\end{eqnarray}
Introducing  again $Y_{c_5}= n_{c_5}\, (\frac{k_b T}{\hbar c})^3$, 
$Y^{(0)}_{c_5}= n^{(0)}_{c_5}\, (\frac{k_b T}{\hbar c})^3$ and 
$\lambda_{c_5} = \frac{<\sigma v>_{c_5}  \, m_{q_5} c^2}{h_m\, (\hbar c)^3}$, with the 
same $x$ and $h_m$ as above, we obtain the equation 
 $\frac{dY_{c_5}}{dx} = 
\frac{\lambda_{c_5}}{x^2}\, (Y^{2}_{q_5} - Y_{c_5} \,Y^{(0)}_{q_5}\, 
\frac{Y^{(0)}_{q_5}}{Y^{(0)}_{c_5}} )$. 

The number density of the fifth family quarks $n_{q_5}$  (
and correspondingly  $Y_{q_5}$), which has above the 
temperature of the binding energy of the clusters of the fifth family quarks (almost) 
reached the decoupled value, starts to decrease again due  to the formation of the clusters of the 
fifth family quarks (and anti-quarks) as well as due to forming the bound state of 
the fifth family quark and an anti-quark, which annihilates into gluons.  
It follows  
\begin{eqnarray}
\label{boltzq2}
a^{-3}\frac{d( a^3 n_{q_5})}{dt} &=& 
< \sigma v>_{c_5}\; n^{(0)}_{q_5}\, n^{(0)}_{q_5} 
\left[ -\left( \frac{n_{q_5}}{n^{(0)}_{q_5}} \right)^2 + \frac{n_{c_5}}{ n^{(0)}_{c_5}} -  
\frac{\eta_{(q\bar{q})_b}}{\eta_{c_5}} \;\left( \frac{n_{q_5}}{n^{(0)}_{q_5}} \right)^2 \right] + \nonumber\\ 
&& < \sigma v>_{q\bar{q}}\; n^{(0)}_{q_5} n^{(0)}_{\bar{q}_5}\,
\left(- \frac{n_{q_5} n_{\bar{q}_5}}{n^{(0)}_{q_5} n^{(0)}_{\bar{q}_5}} + 
 \frac{n_{q} n_{\bar{q}}}{n^{(0)}_{q} n^{(0)}_{\bar{q}}} \right) + \nonumber\\ 
&&< \sigma v>_{gg} \; 
n^{(0)}_{q_5} n^{(0)}_{\bar{q}_5}\,
\left(- \frac{n_{q_5} n_{\bar{q}_5}}{n^{(0)}_{q_5} n^{(0)}_{\bar{q}_5}} +  
\frac{n_{g} n_{g}}{n^{(0)}_{g} n^{(0)}_{g}} \right), 
\end{eqnarray}
with $\eta_{(q\bar{q})_b}$ and $\eta_{c_5}$ defined in Eq.~\ref{sigmasq}.
Introducing the above defined $Y_{q_5}$  and $Y_{c_5}$ the Eq.~\ref{boltzq2} transforms into 
$\frac{dY_{q_5}}{dx} = 
\frac{\lambda_{c_5}}{x^2}\, (- Y^{2}_{q_5} + Y_{c_5} \,Y^{(0)}_{q_5}\, 
\frac{Y^{(0)}_{q_5}}{Y^{(0)}_{c_5}} ) + \frac{\lambda_{(q\bar{q})_b}}{x^2}\, (-  Y^{2}_{q_5})
+ \frac{\lambda_{q_5}}{x^2}\, (Y^{(0)2}_{q_5} - Y^{2}_{q_5})$, 
with $\lambda_{(q\bar{q})_b} = \frac{<\sigma v>_{(q\bar{q})_b}  \, m_{q_5} c^2}{h_m\, (\hbar c)^3}$ 
(and with the same $x$ and $h_m$ as well as $\lambda_{c_5}$ and $\lambda_{q_5}$
as defined above). We solve this equation together with the above equation 
for $Y_{c_5} $. 

Let us look at also at the Thompson scattering of gluons on the bound states, 
destroying clusters, which 
starts to be negligible when the rate for gluons to scatter off the quarks ($n_{q_5} \,\sigma_T \,c$) 
starts to be smaller than the expansion rate ($H= \sqrt{\frac{8\pi^3 \,g^*}{45}} 
\; \frac{(T k_b)^2 \,c}{\hbar c\, m_{pl} c^2}$), with $g^*$ defined above. 
Recognizing that the binding energy of Table~\ref{TableI.} is approximately  $\frac{1}{100}$ the mass 
of the fifth family quarks we get the requirement that the bound states can be formed when 
$n_{q_5} << 3. 10^{-25} (\frac{m_{q_5} \, c^2}{{\rm GeV}})^4$ 
fm$^{-3}$, which for $m_{q_5}\, c^2 = 1$ TeV gives $n_{q_5} << 3. 10^{-13}$ fm$^{-3}$ and  for 
$m_{q_5}\, c^2 = 10$ TeV gives $n_{q_5} << 3. 10^{-9}$ fm$^{-3}$. One can easily check from the 
solutions of the Boltzmann equations that this requirements are fulfilled. 

\begin{figure}[h]
\begin{center}
\includegraphics[width=15cm,angle=0]{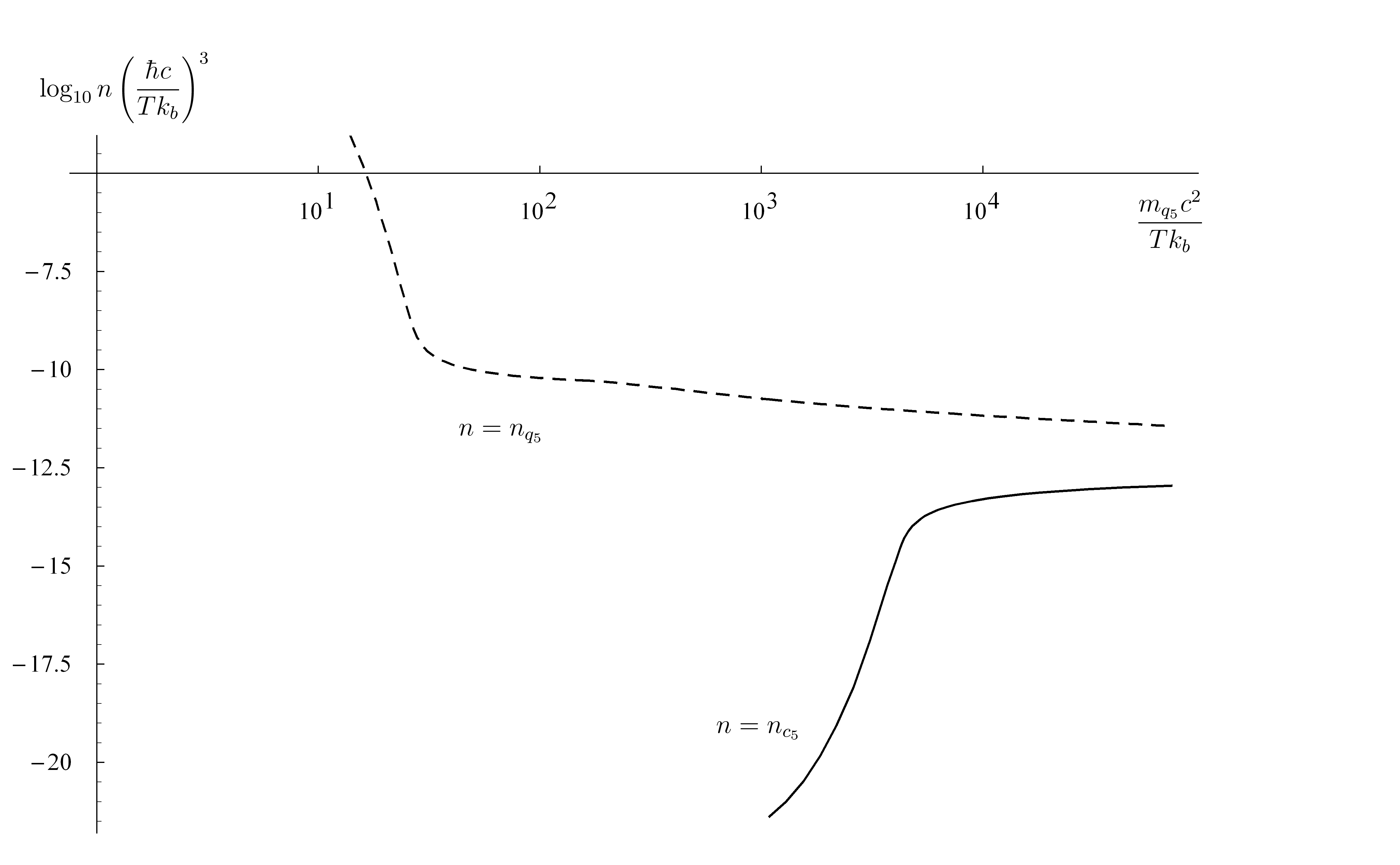}
\caption{The dependence of the two number densities $n_{q_5}$ (of the fifth family quarks) and $n_{c_5}$ (of 
the fifth family clusters) as the function of $\frac{m_{q_5} \, c^2}{T \, k_b}$ is presented 
for the special values $m_{q_5} c^2= 71 \,{\rm TeV}$, $\eta_{c_5} = \frac{1}{50}$ and $\eta_{(q\bar{q})_b}=1$.
We take $g^*=91.5$. }
\end{center}
\label{DiagramI.}
\end{figure}

Solving the two Boltzmann  equations (Eqs.~\ref{boltzq1},~\ref{boltzc}) we obtain 
 the number density of the fifth family quarks  $n_{q_5}$ (and 
anti-quarks) and the number density of the fifth family baryons $n_{c_5}$ (and anti-baryons)
 as a function of the mass of the fifth family quarks and the two parameters $\eta_{c_5}$ 
 and $\eta_{(q\bar{q})_3}$. The evaluations are made, as we explained above, 
within the approximation that the formation of the baryons is determined by the formation of the two 
quark (anti-quarks) clusters, corrected by the parameter  
$\eta_{c_5}$ (Eq.~\ref{sigmasq}). The parameter $\eta_{(q \bar{q})_b}$ corrects the formation of a 
quark---anti-quark pair and its annihilation into gluons. 
In Diagram~\ref{DiagramI.} both number densities (multiplied my $(\frac{\hbar \, c}{T k_b})^3$, which is 
$Y_{q_5}$ and $Y_{c_5}$, respectively for the quarks and the clusters of quarks) as a function 
of  $ \frac{m_{q_5} \, c^2}{T k_b}$ for $\eta_{(q\bar{q})_3}=1$ and $\eta_{c_5}=\frac{1}{50}$ are presented. 
The calculation is performed up to $T k_b=1$ GeV.

 The quarks and anti-quarks are at high temperature 
($\frac{m_{q_5} c^2}{T k_b}<< 1$) in thermal equilibrium with the plasma (as are also all the other 
families and bosons of  lower masses). 
As the temperature of the plasma (due to the expansion of the universe) drops close to the 
mass of the fifth family 
quarks, quarks and anti-quarks scatter into all the other (ultra) relativistic fermions and bosons.  
At the temperature close to  the binding energy of the quarks in a cluster, the clusters of 
baryons start to be formed. We evaluated the number density  $n_{q_5} (T) \,
(\frac{\hbar c}{T k_b})^3 = Y_{q_5} $  
of the fifth family quarks (and anti-quarks) and the number density of the fifth family baryons 
$n_{c_5} (T) \,(\frac{\hbar c}{T k_b})^3 = Y_{c_5} $ for several choices  of  
$m_{q_5}, \eta_{c_5}$ and $\eta_{(q \bar{q})_b}$ up to $T k_b = 1$ GeV $=\frac{m_{q_5} c^2}{x} $.  

>From the calculated decoupled number density of baryons and anti-baryons of the fifth family quarks  
(and anti-quarks) $n_{c_5}(T_1)$ at temperature $T_1 k_b=1$ GeV, where we stopped our 
calculations  as a function of the quark mass and of the 
two parameters $\eta_{c_5}$ and $\eta_{(q\bar{q})_b}$, the today's mass density of the dark matter 
follows 
\begin{eqnarray}
\label{dm}
\rho_{dm} &=& \Omega_{dm} \rho_{cr}= 2 \, m_{c_5}\, n_{c_5}(T_1) \, 
\left(\frac{T_0}{T_1} \right)^3 \frac{g^*(T_1)}{g^*(T_0)},
\end{eqnarray}
with $T_0 = 2.5 \,\cdot 10^{-4}\,\frac{{\rm eV}}{k_b}$, $g^*(T_0)= 2 + \frac{7}{8}\,\cdot3 \,\cdot \,
(\frac{4}{11})^{4/3}$, $g^*(T_1)= 2 + 2\,\cdot 8 + \frac{7}{8}\, (5\cdot3\cdot2\cdot2 + 6\cdot 
2\cdot2)$ and  
 $\rho_{cr} \,c^2 \,\approx \frac{3\, H^{2}_{0}\, c^2}{8 \pi G} 
 \approx 5.7\, \cdot 10^3 \frac{{\rm eV}}{cm^3}$, factor $2$ counts 
 baryons and anti-baryons (since the spin  of baryons is taken into account in $n_{c_5}$).

The influence of the choice of the parameters $\eta_{c_5}$ and $\eta_{(q \bar{q})_b}$  on the number 
of the fifth family baryons and anti-baryons (together with the evaluation of the numerical errors) 
is used as a measure for the accuracy with which we  evaluated the fifth family mass.

\begin{table}
\begin{center} \begin{tabular}{|c||c|c|c|c|}
\hline
$\frac{m_{q_5} c^2}{{\rm TeV}}$&$\eta_{(q\bar{q})_b}=1$&$\eta_{(q\bar{q})_b}=\frac{1}{3}$& 
$\eta_{(q\bar{q})_b}=3$&$\eta_{(q\bar{q})_b}=10$\\
\hline\hline
$\eta_{c_5}=1$ & 19 & 11 & 37& \\
\hline
$\eta_{c_5}=3$ & 15 & 9.5 & 27& \\
\hline
$\eta_{c_5}=\frac{1}{3}$ & 25 & 14 & 54& \\
\hline
$\eta_{c_5}=10 $& 13 & -- & 22&\\
\hline
$\eta_{c_5}= \frac{1}{10}$ & 39 & 20 & 84& \\
\hline
$\eta_{c_5}=\frac{1}{50}$ & 71 & -- &-- &  417\\
\hline
\end{tabular}
\end{center}
\caption{\label{TableII.} The fifth family quark mass is presented, calculated for different 
choices of $\eta_{c_5}$ (which takes care of the probability that a colourless cluster of three 
quarks (anti-quarks) instead of two are formed) and of $\eta_{(q\bar{q})_b}$ (which takes care of the 
annihilation of a bound state of quark---anti-quark) from Eqs.~(\ref{dm}, \ref{boltzc}, \ref{boltzq1}). }
\end{table}

We read from Table~\ref{TableII.} the mass interval for the fifth family quarks' mass, 
which fits Eqs.~(\ref{dm}, \ref{boltzc}, \ref{boltzq1}):
\begin{eqnarray}
\label{massinterval}
10 \;\; {\rm TeV} < m_{q_5}\, c^2 < {\rm a\, few} \cdot 10^2 {\rm TeV}.
\end{eqnarray}
From this mass interval we estimate from Table~\ref{TableI.} the cross section for the 
fifth family neutrons $\pi (r_{c_5})^2$:
\begin{eqnarray}
\label{sigma}
10^{-8} {\rm fm}^2 \, < \sigma_{c_5} < \, 10^{-6} {\rm fm}^2.
\end{eqnarray}
(It is  at least $10^{-6} $ smaller than the cross section for the first family neutrons.)

\section{ Concluding remarks}
\label{conclusion}

We estimated in this paper the possibility that a new  stable  family, 
predicted by the approach unifying spin and 
charges~\cite{pn06,n92,gmdn07},  
forms baryons which are the dark matter constituents. The approach (proposed by S.N.M.B.)
 is to our knowledge the only proposal  
in the literature so far which offers the mechanism for generating families, 
if we do not count those which on one or another way just assume more 
than three families. 
We evaluated the limits on the properties of the stable fifth family quarks due to the 
cosmological observations and the direct experiments.

We use the simple hydrogen-like model to evaluate the 
properties of these heavy baryons and their interaction 
among themselves and with the ordinary  nuclei. We take into account that for masses of the order 
of $1$ TeV/$c^2$ or larger the one gluon exchange determines the force among the constituents of 
the fifth family baryons. Studying the interaction of these baryons with the ordinary matter we 
take into account  that 
the weak interaction starts to dominate over the 
''nuclear interaction'' which the fifth family neutron manifests 
for massive enough clusters ($m_{q_5}> 10^4$ TeV), while  the non relativistic 
fifth family baryons interact among themselves with the weak force only.

We assume 
that in the evolution of our universe $q_5$ and $\bar{q}_5$ were formed with no asymmetry. We study  
the freeze out procedure of the fifth family quarks and anti-quarks and the formation of  
baryons and anti-baryons up to the temperature  $T k_b = 1$ GeV,  when the colour phase transition 
starts which depletes all the fifth family quarks and anti-quarks while the colourless
fifth family neutrons with very small scattering cross section decouples long before (at $T k_b = 100$ GeV).

While the measured density of  the  dark matter 
does not put much limitation on the properties of heavy enough clusters,  
the DAMA experiments~\cite{rita0708} limit (provided that they measure 
our heavy fifth family clusters) the quark mass 
to:  $ 200 \,{\rm TeV} < m_{q_{5}}c^2 < 10^5\, {\rm TeV}$.   
The estimated cross section for the dark matter cluster to 
(elastically, coherently and nonrelativisically) scatter on the (first family) nucleus is in this case 
determined on the lower mass limit by the ''nuclear force'' of the fifth family clusters 
($ (3\cdot 10^{-5}\,A^2\, {\rm fm} )^2$) 
and on the higher mass limit by the weak force 
($ ( A (A-Z)\, 10^{-6} \, {\rm fm} )^2 $). 

The cosmological evolution 
suggests for the mass limits the range $10$ TeV $< m_{q_5} \, c^2 < {\rm a \, few} \cdot 10^2$ TeV 
and for the  scattering cross sections 
$ 10^{-8}\, {\rm fm}^2\, < \sigma_{c_5}\, <   10^{-6} \,{\rm fm}^2  $. 
Accordingly we conclude that if the DAMA experiments are measuring our fifth family neutrons,  
the mass of the fifth family quarks is a few hundred  TeV $/c^2$.

In the ref.~\cite{mbb}~\footnote{ The referee of PRL suggested 
that we should comment on the paper~\cite{mbb}.} the authors 
study the limits on a scattering cross section of 
a heavy dark matter cluster with the ordinary matter. They assume (approximately) the same number of 
particles and antiparticles in the dark matter. They treat the conditions under which 
would  
the heat flow  which would follow 
from the annihilation of dark matter  particles and anti-particles in the Earth core start to be noticable. 
Using their limits we conclude that our fifth family baryons of the mass of a few hundreds TeV/${c^2} $ 
have  for a factor more than $100$ too small scattering amplitude with the ordinary matter to cause a measurable 
heat flux.


Our rough estimations predict that, if the DAMA experiments 
observe the events due to our (any) 
heavy family members, (or any heavy enough family cluster with 
small enough cross section),  
the CDMS experiments~\cite{cdms} will in the near future observe  
a few events as well. 
%
If CDMS will not confirm the heavy family events, then we must conclude, 
trusting the DAMA experiments, that either our 
fifth family clusters have much higher cross section due to the possibility that $u_5$ is lighter than 
$d_5$  so that their velocity slows down when 
scattering on nuclei of the earth above the measuring apparatus 
bellow the threshold of the CDMS experiment (and that there must be in this case the fifth family 
quarks---anti-quarks asymmetry) 
while the DAMA experiment still observes them, 
%
or the fifth family clusters (any heavy stable family clusters) are not what forms the dark matter.

Let us comment again the question whether there is at all possible (due to electroweak experimental
data) that there exist more than three up to now observed families, that is, whether the approach 
unifying spin and charges  by predicting the fourth and the stable fifth  
family (with neutrinos included) contradict the observations. In the 
ref.~\cite{mdnbled06} the properties 
of all the members of the fourth family were studied (for  one particular choice of breaking the starting 
symmetry). The predicted fourth family neutrino mass is at around $100$ GeV/$c^2$ or higher, therefore it 
does not due to the detailed analyses of the electroweak data done by the Russian group~\cite{okun} 
contradict any experimental data. (
The stable fifth family neutrino has due to our calculations 
considerably higher mass. Accordingly none of these two neutrinos contradict  
the electroweak data. They also do not 
contradict the nucleosynthesis, since to the nucleosynthesis only the neutrinos with masses 
bellow the electron mass contribute. 
The fact that the fifth family baryons might form the dark matter does not contradict  
 the measured (first family) baryon number and its ratio to the photon 
 energy density as well, as long as the fifth family quarks are 
 heavy enough ($>$1 TeV). All the measurements, which connect the baryon and the photon 
 energy density, relate to the moment(s) in the history of 
 the universe, when the baryons (of the first family) where formed ($m_1 c^2 \approx 
  k_b T = 1$ GeV and lower)  and the electrons and nuclei were  forming  atoms ($k_b \,T 
 \approx 1$ eV). The chargeless (with respect to the colour and electromagnetic 
 charges, not with respect to the weak charge) clusters of the fifth family were 
 formed long before (at $T k_b \approx E_{c_5}$ (Table~\ref{TableI.})). They  manifest 
 after decoupling from the plasma (with their small number density and  small cross 
 section) (almost) only their gravitational  interaction.


Let us conclude this paper with the recognition:   
 If the approach unifying spin and charges is the right way beyond the 
 standard model of the electroweak and colour interaction,  
 then more than three 
 families of quarks and leptons do exist, and the stable 
 (with respect to the age of the universe) fifth family of quarks and leptons 
 is the candidate to form the dark matter.

 \section{ Appendix I. Three fifth family quarks' bound states}
 \label{betterhf}

 We look for the ground  state solution of the Hamilton equation  $H\,
 |\psi\rangle= E_{c_5}\,|\psi\rangle $ for a cluster of three heavy quarks with 
 \begin{eqnarray}
  H=\sum_{i=1}^3 \,\frac{p_{i}^2}{2 \,m_{q_5}} 
  -\frac{2}{3}\,  \, \sum_{i<j=1}^3
   \frac{\hbar c \; \alpha_c}{|\vec{x}_i-\vec{x}_j|},   
 \end{eqnarray}
 in the center of mass motion
 \begin{eqnarray}
  \vec{x}=\vec{x}_2 - \vec{x}_1,\quad  \vec{y}=\vec{x}_3 - \frac{\vec{x}_1+\vec{x}_2}{2},\quad
  \vec{R}=\frac{\vec{x}_1+\vec{x}_2+\vec{x}_3}{3}, 
 \end{eqnarray}
 assuming the antisymmetric colour part  ($|\psi\rangle_{c,\, \cal{A} }$), 
 symmetric spin and weak charge part  ($|\psi\rangle_{w  \, {\rm spin},\, \cal{S}  }$)
 and symmetric space part ($|\psi\rangle_{{\rm space}, \, \cal{S}}$). 
 For the space part we take the hydrogen-like wave functions 
 $ \psi_a(\vec{x})=$$\frac{1}{\sqrt{\pi a^3}} \; e^{-|\vec{x}|/a}$ and 
  $\psi_b(\vec{y})=$$\frac{1}{\sqrt{\pi b^3}} \; e^{-|\vec{y}|/b}$, allowing $a$ and $b$ to 
  adapt variationally.
  Accordingly $\langle\, \vec{x}_1, \vec{x}_2, \vec{x}_3|\psi\rangle_{{\rm space}\, \cal{S}}=
  \mathcal{N}
 \left( \psi_a(\vec{x}) \psi_{b}(\vec{y}) + \textrm{symmetric permutations} \right)$. It follows 
 $\langle\, \vec{x}_1, \vec{x}_2, \vec{x}_3|\psi\rangle_{{\rm space}\, \cal{S}}=
 \mathcal{N} \,  \left(  2 \psi_a(\vec{x}) \psi_{b}(\vec{y})   +
  2 \psi_a(\vec{y}-\frac{\vec{x}}{2}) \psi_{b}(\frac{\vec{y}}{2}+\frac{3 \vec{x}}{4}))  +
 2 \psi_a(\vec{y}+\frac{\vec{x}}{2}) \psi_{b}(\frac{\vec{y}}{2}-\frac{3 \vec{x}}{4}) \right) $.
 
 The Hamiltonian in the center of mass motion reads 
 $H=\frac{p_x^2}{2 (\frac{m_{q_5}}{2})}+\frac{p_y^2}{2 (\frac{2m_{q_5}}{3})}+\frac{p_R^2}{2 
 \cdot 3 m_{q_5}}
 -\frac{2}{3} \hbar c \; \alpha_c \left(\frac{1}{x}+\frac{1}{|\vec{y}+\frac{\vec{x}}{2}|}+
 \frac{1}{|\vec{y}-\frac{\vec{x}}{2}|} \right).
 $
 Varying the expectation value of the Hamiltonian with respect to $a$ and $b$ 
 it follows: $\frac{a}{b}=1.03, \, \frac{a\, \alpha_c\, m_{q_5}\, c^2}{\hbar c} = 1.6$. 
 
 Accordingly we get for the binding energy  $ E_{c_5}=0.66\; m_{q_5}\, c^2 \alpha_{c}^2$ and for the size 
 of the cluster $\sqrt{\langle |\vec{x}_2-\vec{x}_1|^2 \rangle} = 2.5\, \frac{\hbar c}{\alpha_c m_{q_5 \, c^2}}
 $.

 To estimate  the  mass difference between $u_5$ and $d_5$
 for which $u_5 d_5 d_5$ is stable we treat the electromagnetic ($\alpha_{elm}$) and weak ($\alpha_w $) 
 interaction as a small correction 
 to the above calculated binding energy: $H'=  \alpha_{elm\,w}  \; \hbar c \,
 \left(\frac{1}{x}+\frac{1}{|\vec{y}+\frac{\vec{x}}{2}|}+
 \frac{1}{|\vec{y}-\frac{\vec{x}}{2}|} \right)$. $\alpha_{elm\,w} $ stays for electromagnetic and 
 weak coupling constants.
 For $m_{q_5}= 200$ TeV we take $\alpha_{elm\,w} =\frac{1}{100}$, then 
 $|m_{u_5}- m_{d_5}|< \frac{1}{3}\,  E_{c_5} \frac{(\frac{3}{2} \alpha_{elm\,w})^2}{\alpha_c^2}
 = 0.5\,\cdot 10^{-4} \; m_{q_5}\,  c^2 $.

 %
%

%
\section*{Acknowledgments} 

The authors  would like to thank  all the participants 
of the   workshops entitled 
"What comes beyond the Standard models", 
taking place  at Bled annually (usually) in  July, starting in 1998, and in particular  
H. B. Nielsen, since all the open problems were there very openly discussed.



\begin{thebibliography}{99}

%
\bibitem{rita0708} R. Bernabei, P. Belli, F. Cappella, R. Cerulli, F. Montecchia, F. Nozzoli, 
A. Incicchitti, D. Prosperi, C.J. Dai, H.H. Kuang, J.M. Ma, Z.P. Ye, 
''Dark Matter particles in the galactic halo: results and implications from DAMA/NaI'', 
 Int. J. Mod. Phys. D13 (2004) 2127-2160, astro-ph/0501412, astro-ph/0804.2738v1.
 %
\bibitem{cdms} Z. Ahmed et al., A Search for WIMPs with the First Five-Tower Data from CDMS, 
pre-print, astro-ph/0802.3530.
%
\bibitem{dodelson} S. Dodelson, Modern Cosmology, Academic Press Elsevier 2003.
%
%
\bibitem{pdg} C. Amsler et al., Phys. Lett. B 667, 1 (2008). 

\bibitem{pn06} A. Bor\v stnik Bra\v ci\v c, N.S. Manko\v c Bor\v stnik, ''On the origin of 
families of fermions and their mass matrices'', 
hep-ph/0512062,  
Phys. Rev. {\bf D 74} (2006) 073013-16. 
%
\bibitem{n92} N.S. Manko\v c Bor\v stnik, ``Spin connection as a superpartner of a vielbein''
Phys. Lett. {\bf B 292} (1992) 25-29,
''Unification of spin and charges'',  
Int. J. Theor. Phys. {\bf 40} (2001) 315-337,  
%
''From the starting Lagrange density to the 
 effective fields for spinors in the Approach unifying spin and charges'' in the Proceedings.
hep-ph/0711.4681, 94-113. 
%
\bibitem{n93} N.S. Manko\v c Bor\v stnik, 
``Spinor and vector representations in four dimensional Grassmann space'', 
J. Math. Phys. {\bf 34} (1993) 3731-3745, ''Unification of spin and charegs in Grassmann space?'',
Mod. Phys. Lett. {\bf A 10}  (1995) 587-595.  
%
\bibitem{gmdn07} G. Bregar, M. Breskvar, D. Lukman, N.S. Manko\v c Bor\v stnik, 
''Predictions for four families by the Approach unifying spin and charges'', 
 New J. of Phys. {\bf 10} (2008) 093002. 
%
\bibitem{hn02hn03} N.S. Manko\v c Bor\v stnik, H.B. Nielsen, ''How to generate
              spinor representations in any dimension in terms of projection operators'', 
              ''How to generate families of spinor'',
J. of Math. Phys. {\bf 43} (2002), (5782-5803), J. of Math. Phys. 
{\bf 44} (2003) 4817-4827. 
%
\bibitem{gnBled07} G. Bregar, N.S. Manko\v c Bor\v stnik, arXiv:0812.0510 [hep-ph], 2-12. 
%
%
%
 %
%
\bibitem{Ruhula} A. De Rujula, Phys. Rev.  D 12 (1975) 147.
%
\bibitem{gmn08} G. Bregar, M. Khlopov, N.S. Manko\v c Bor\v stnik, work in progress. 
%
%
\bibitem{privatecommRBJF} 
We thank cordialy R. Bernabei (in particular) 
and J. Filippini for very informative discussions by emails and in private communication.
%
\bibitem{okun} V.A. Novikov, L.B. Okun,A.N. Royanov, M.I. Vysotsky, ``Extra generations and
discrepancies of electroweak precision data'', 
Phys. Lett. B 529 (2002) 111, hep-ph/0111028, private discussions.
%
%
%
%
%
\bibitem{mdnbled06} M. Breskvar, D. Lukman, N.S. Manko\v c Bor\v stnik,  hep-ph/0612250, p.25-50.
%
\bibitem{mbb} G.D. Mack, J.F. Beacom , G. Bertone, Phys. Rev. D 76 (2007) 043523.

%
\end{thebibliography}
\end{document}